\documentclass{article}

\usepackage{hyperref}
\usepackage[english]{babel}
\usepackage{amsfonts}
\usepackage{amsmath}
\usepackage{mathtools}
\usepackage{amssymb}
\usepackage{epsfig}
\usepackage{alltt}
\usepackage[latin1]{inputenc}
\usepackage{fixme}
\usepackage[table,xcdraw]{xcolor}
\usepackage{pstricks}
\usepackage{pst-node}

\usepackage{enumerate}
\usepackage[T1]{fontenc}
\usepackage[latin1]{inputenc}
\usepackage{amssymb,amsmath}
\usepackage{color}
\usepackage{hyperref}
\usepackage{graphics}

\newcounter{cdobleimpl}
\def\thecdobleimpl{\ifnum\value{cdobleimpl}=1 $\Longrightarrow$:\ \else $\Longleftarrow$:\ \fi}

\newenvironment{proof}{\par\textbf{\textit{Proof.}}\ }{\qed}

\def\squareforqed{\hbox{\rlap{$\sqcap$}$\sqcup$}}
\def\qed{\ifmmode\squareforqed\else{\unskip\nobreak\hfil
\penalty50\hskip1em\null\nobreak\hfil\squareforqed
\parfillskip=0pt\finalhyphendemerits=0\endgraf}\fi}

\newtheorem{theorem}{\bfseries Theorem}%[chapter]
\newtheorem{definition}{\bfseries Definition}%[chapter]
\newtheorem{proposition}{\bfseries Proposition}%[chapter]
\newtheorem{corollary}{\bfseries Corollary}%[chapter]
\newtheorem{lemma}{\bfseries Lemma}%[chapter]
\newtheorem{example}{\bfseries Example}%[chapter]

\newcommand{\ismcomment}[1]{}
\newcommand{\bdfn}{\begin{definition} \begin{rm}}
\newcommand{\edfn}{\end{rm}$ $\qed \end{definition}}
\newcommand{\bthm}{\begin{theorem} \begin{rm}}
\newcommand{\ethm}{\end{rm}$ $\qed \end{theorem}}
\newcommand{\bprop}{\begin{proposition} \begin{rm}}
\newcommand{\eprop}{\end{rm}\qed\end{proposition}}
\newcommand{\bcor}{\begin{corollary}\begin{rm}}
\newcommand{\ecor}{\end{rm} \end{corollary}}
\newcommand{\blem}{\begin{lemma} \begin{rm}}
\newcommand{\elem}{\end{rm}\qed\end{lemma}}
\newcommand{\bfact}{\begin{fact} \begin{rm}}
\newcommand{\efact}{\end{rm} \end{fact}}
\newcommand{\bex}{\begin{example} \begin{rm}}
\newcommand{\eex}{\end{rm}$ $\qed  \end{example}}

\newcommand{\bprf}{\begin{proof}}
\newcommand{\eprf}{\end{proof}}

\newenvironment{sketch}%
{\nopagebreak[4]\vspace*{0.2em}\noindent{\bf\it Proof Sketch}:\hspace{1ex}}
  {}%{\nopagebreak\hfill$\Box$}
\newcommand{\bprfsketch}{\begin{sketch}}
\newcommand{\eprfsketch}{\end{sketch}}

\newcommand{\comen}[1]{}

\newcommand{\letbar}[1]{\mbox{I\kern-0.23em#1}}
\newcommand{\nat}{\mathbb{N}}
\newcommand{\real}{\mathbb{R}}

\newcommand{\calA}{{\cal A}}

\newcommand{\calU}{{\cal U}}

\newbox\arriba
\newbox\abajo
\newbox\CaracterInterno
\newbox\CaracterDerecha
\newdimen\anchura
\def\MacrosTranGeneral#1#2#3#4#5#6{%
  \setbox\CaracterInterno=\hbox{\mathsurround=0pt$\mathord#4$}
  \setbox\CaracterDerecha=\hbox{\mathsurround=0pt$\mathord#3$}
  \setbox\arriba=\hbox{$#1#2$}
  \setbox\abajo=\hbox{\mathsurround=0pt%
                      \anchura=\wd\arriba%
                      \advance \anchura by 0.5em%
                      \divide \anchura by \wd\CaracterInterno%
                      \multiply \anchura by \wd\CaracterInterno%
                      \copy\CaracterInterno\kern\SeparacionInternaFlecha
                      \hbox to \anchura{%
                          $\cleaders%
                            \hbox{\kern\SeparacionInternaFlecha\copy\CaracterInterno}
                            \hfill$}%
                      \kern\SeparacionExternaFlecha\copy\CaracterDerecha}
  \mathrel{{\buildrel\vbox{\copy\arriba \kern\SeparacionFlechaArriba} %
    \over{\copy\abajo^{#6}}}_{#5}}
  }
\def\MacrosTranGeneralProp#1#2#3#4#5{\mathchoice%
  {\MacrosTranGeneral{\scriptstyle}{#1}{#2}{#3}{#4}{#5}}
  {\MacrosTranGeneral{\scriptstyle}{#1}{#2}{#3}{#4}{#5}}
  {\MacrosTranGeneral{\scriptscriptstyle}{#1}{#2}{#3}{#4}{#5}}
  {\MacrosTranGeneral{\scriptscriptstyle}{#1}{#2}{#3}{#4}{#5}}}

\def\MacrosNoTran#1{%
  \def\SeparacionInternaFlecha{-0.3em}
  \def\SeparacionExternaFlecha{-0.5em}
  \def\SeparacionFlechaArriba{-3pt}
  \MacrosTranGeneralProp{#1\kern 0.5em}{{\not\rightarrow}}{-}{}{}}

\title{\LARGE \bf
Measuring the benefits of lying in MARA under egalitarian social welfare\thanks{This paper was published in 2020 IEEE International Conference on Systems, Man, and Cybernetics (SMC).
The present version is the author's accepted manuscript. This work has been partially supported by projects TIN2015-67522-C3-
3-R, PID2019-108528RB-C22, and by Comunidad de Madrid as part of the program S2018/TCS-4339 (BLOQUES-CM) co-funded by EIE Funds of the European Union.}}

\author{Jonathan Carrero$^1$, Ismael Rodr{\'\i}guez$^{1,2}$,
        and Fernando Rubio$^{1,2}$
\thanks{$^{1}$Dpto. Sistemas Inform{\'a}ticos y Computaci{\'o}n, Facultad de Inform{\'a}tica, Universidad Complutense de Madrid, 28040 Madrid, Spain.
        {\tt\small joncarre@ucm.es, isrodrig@sip.ucm.es, fernando@sip.ucm.es}}
\thanks{$^{2}$Instituto de Tecnolog{\'\i}as del Conocimiento, Universidad Complutense de Madrid, 28040 Madrid, Spain.
}}

\begin{document}

\date{}

\maketitle

% --- ABSTRACT
\begin{abstract}
When some resources are to be distributed among a set of
agents following egalitarian social welfare, the goal is
to maximize the utility of the agent whose utility turns out
to be minimal. In this context, agents can have an incentive to lie about their actual preferences, so that more valuable resources are assigned to them. In this paper we analyze this situation, and we present a practical study where genetic algorithms are used to assess the benefits of lying under different situations.
\end{abstract}


\section{Introduction}
Computer Science and Economics have been successfully applied to each other for decades. 
%On the one hand, computational environments have dramatically changed economic interactions by means of, for instance, e-commerce, collaborative economies, or virtual coins. %~\cite{ecommerce01,bitcoin08,sharingEco17}. 
%On the other hand, 
Economic concepts and mechanisms can be applied to deliver scarce resources %such as e.g. processing time, bandwidth, or memory 
in a computational environment (see e.g.~\cite{grid05,leon10}),
%\cite{mariposa96,grid05,leon10}
%Economic notions have even been proposed as the cornerstone for alternative programming paradigms~\cite{agoric88}, 
while
%In particular, 
Computer Science has %even 
been applied to identify the difficulty of developing several economic tasks (see e.g.~\cite{marketing16,rrr17,chen2018oxford,monaco19}). %, as well as to find the  efficiency limits of the human markets themselves~\cite{marketsPNP11}. 

Actually, some of the challenges one finds in the problem of delivering resources are the same regardless of the nature of agents (natural or artificial). Multiagent Resource Allocation (MARA) deals with the problem of delivering some resources to some agents, given their preferences over the resources. The definition of a particular MARA environment depends on many factors, and the first decision to be addressed is what the goal of the distribution is. For instance, the goal may be maximizing the revenue of the particular agent making the distribution (or {\it auctioneer}) by means of the payments done by bidders for the resources or bundles of resources (called {\it Winner Determination Problem}~\cite{winnerDet06} in the latter case). Alternatively, if the goal is satisfying the preferences of the agents these resources are assigned to, other alternatives unfold depending on the kind of society satisfaction ({\it social welfare}) to be pursued. For instance, the goal may be maximizing the addition of utilities of all agents these resources are assigned to, which is the so-called {\it utilitarian} social welfare; or it may consist in maximizing the utility of the agent whose utility turns out to be minimal, which is called {\it egalitarian} social welfare (think, for instance, in distributing humanitarian resources after a disaster); or other criteria may be considered (see e.g.~\cite{socialReview06}). Regarding how agent preferences are denoted, agents may just assign a numerical value to each available resource and assume {\it additive} preferences (i.e. the utility of a bundle of resources is the addition of the utilities of all resources in the bundle). Alternatively, non-additive settings where bundles of resources can be given utilities different to the simple addition (higher or lower) can be denoted in several ways~\cite{nnr14}. Regarding the process followed to achieve the desired delivery, we may assume that it is performed in one-shot by some party (say an auctioneer) that initially receives the preferences of all agents, or that the desired delivery is the result of iterating local interactions between agents according to some local rules~\cite{ems11}.

Let us consider deliveries performed in one shot. Note that agents may have an incentive to communicate false preferences to the auctioneer in order to maximize their utility. For instance, if the utilitarian social welfare is the goal of the delivery, then agents may benefit by pretending the utility they get for each resource is higher than it is, as the addition of utilities of agents is maximized by giving resources to those agents wanting them the most. In order to cope with this problem, the Generalized Vickrey Auction (GVA)~\cite{VickreyGeneral99} introduces a payment mechanism to eliminate any incentive for communicating preferences different to the true ones. In this auction, the distribution of resources maximizing the addition of utilities is found and assigned to agents, but in addition each agent has to pay an amount equivalent to the {\it loss of utility} of the other agents due to the participation of the agent. More specifically, this payment is the subtraction of the utility of the others in a distribution where the agent did not exist, {\it minus} the utility of the others according to the actual distribution of resources.
After taking this payment into account, the final utility of each agent turns out to be the addition of utility of {\it all} agents (due to the resources the agent receives and, on the other side, the payment {\it reduction} due to the utility of the other agents in the actual distribution) {\it minus} another term the agent has no control at: the payment due to the distribution that would be reached if the agent did not participate. Thus, to the extent that can be controlled by the agent, the utility of the agent (due to the resources it receives and due to the payment) is maximized by maximizing the addition of utility of {\it all} agents, which is actually maximized by communicating the true preferences. This property, usually called {\it strategy-proofness}, eliminates the incentive for strategic behavior in the GVA ---although this mechanism has other drawbacks (see e.g.~\cite{VickreyMalo96,VickreyMalo07}, where problems related with cheating, disclosure of confidential information, etc. are discussed).

To the best of our knowledge, no similar ``falsehood disincentivizing'' mechanism has been found for the case of the egalitarian social welfare. In fact, the intrinsic discontinuities of the function to be maximized in this case hinder this goal, as the utility of the agent with minimum utility dramatically changes by introducing slight resource allocation changes. Apparently, we could hardly make the utility of agents coincide with the egalitarian social welfare by introducing (rational, not extremely distorting) payments, like GVA does with the utilitarian social welfare. Note that not disincentivizing the falsehood is a problem indeed: if the distribution mechanism just delivers the resources in such a way the egalitarian social welfare is maximized, then agents do have a trivial incentive for lying. In fact, an agent just has to do the opposite as we said for the utilitarian social welfare: by falsely {\it reducing} the values of the utilities given by resources, an agent increases its chances to be the agent with minimum utility, and in turn the agent whose utility is to be maximized by the distribution. Thus, agents can trivially increase their profit by underestimating their preferences. 

Despite the apparent impossibility to fully eliminate the incentive for lying (i.e. sending false preferences) in this kind of distribution, it is easy to notice that some simple rule changes can make it harder to get profit by lying or, at least, to {\it certainly} get profit by lying. For instance, if all agents are forced to make the addition of their utilities for individual resources be equal to some fix constant, then agents cannot arbitrarily underestimate their preferences for {\it all} resources, as underestimating some resource implies the risky action of overestimating another. Even if we cannot {\it fully} disincentivize false preferences under the egalitarian social welfare, we can still measure to what extent some modifications of the rules do it ---and check out if some of them virtually disincentivize false preferences {\it in practice}. Note that some ways of lying could have a higher probability of being profitable than others. Moreover, let us assume the agent willing to lie has an estimation of the preferences the other agents will communicate (regardless of whether they are true or not). Then, the optimal set of {\it false preferences} for this agent is the set such that, if it is communicated to the auctioneer {\it and} the other agents also send the preferences estimated by this agent, then the distribution of resources maximizing the egalitarian social welfare will also maximize its {\it true} utility ---maybe beyond the utility achieved by sending the true preferences. Note that this so-called optimal set of false preferences could not be so good if the preferences sent by the other agents turn out to be different to those estimated by the agent. Actually, some deviation (i.e. estimation imprecision) is expected in any real setting, so the suitability of that alleged optimal set of false preferences might strongly depend on the degree of that deviation: maybe using these false preferences is better than saying the truth if that deviation is low, but it is not if the deviation is high. Moreover, some distribution mechanism rules (e.g. forcing the addition of utilities to be equal to some constant) may make the distributions for these optimal false preferences be more {\it sensitive} to those deviations ---to the extent that, even under small and likely deviations, saying the truth is better than lying in general. In this case, we could say that the incentive for lying is {\it practically} eliminated under those particular rules. 

The goal of this paper will be investigating, by means of experiments, if we can disincentive  lying in practice in the context of egalitarian social welfare. Experiments will be conducted where some agent lies according to some predefined lying strategies or by using the lie that would be optimal provided that the other agents behaved exactly as estimated by the agent.    
%which rules modifications bring a higher disincentive for lying, i.e. which ones make the optimal sets of false preferences be less robust to deviations in the estimated preferences. {\bf Adaptar esto segun lo que realmente hagamos:} 
We will assume a basic egalitarian social welfare MARA setting where the preferences are additive, although our experiments could be easily extended to consider other more complex preferences. 
Actually, as maximizing the egalitarian social welfare is NP-hard, 
%and hard to approximate even in this preferences setting, so 
solving each instance of this optimization problem will require approximate algorithms (in particular, we will choose genetic algorithms, although any other metaheuristic could be used, for instance~\cite{xing2014innovative,RRR17JoCS,RR19water}). %,molina2020comprehensive}).
In a nutshell, our experiments will show that, if the addition of preferences is required to be equal to some given constant, then (a) no predefined lying strategy is profitable; and (b) using the optimal lie according to some estimation of the preferences of the other agents will be worse than saying the truth when that estimation is even slightly imprecise.  

The rest of the paper is organized as follows. In the next section we formally introduce the problems under consideration. Section~\ref{sec:experiments} reports our experimental results, and conclusions %and future works lines 
are given in Section~\ref{sec:conclusions}.

\section{Formal model}

In this section  we introduce the formal notions and problems under consideration in this paper.

\bdfn Let $R=\{r_1,\ldots,r_m\}$ be a set of resources and $A=\{A_1,\ldots,A_n\}$ be a set of agents.

The set of possible allocations of $R$ to $A$ is the set $\calA=\underbrace{A\times \ldots \times A}_{m}$. Given $\alpha\in\calA$ with $\alpha=(\alpha_1,\ldots,\alpha_m)$, we say that the allocation $\alpha$ assigns each resource $r_j$ to agent $\alpha_j$ for all $1\leq j\leq m$.\edfn

A utility function will be a function receiving a distribution of resources and returning a real number. As usual, utility functions will be used to denote the preferences of agents for bundles of resources: if the utility function of an agent is $u$ and $u(\alpha_1)>u(\alpha_2)$, then the agent prefers distribution $\alpha_1$ over distribution $\alpha_2$.

\bdfn A utility function is a function \hbox{$u : \calA \longrightarrow \real^{\geq 0}$}.

We say that $u$ {\it depends only} on agent $A_i$ if, for all $\alpha=(\alpha_1,\ldots,\alpha_m)\in\calA$ and $\beta=(\beta_1,\ldots,\beta_m)\in\calA$ fulfilling  $\alpha_j = \beta_j = A_i$ or $\alpha_j \neq A_i \neq \beta_j$ for each $1\leq j\leq m$, we have $u(\alpha) = u(\beta)$.
In addition, we also say that $u$ is {\it additive} for agent $A_i$ if for all $\alpha=(\alpha_1,\ldots,\alpha_m)\in\calA$ we have $u(\alpha)=\sum_{\{j\;|\;\alpha_j = A_i\}} u(\underbrace{A_k,\ldots,A_k}_{i-1},A_i,\underbrace{A_k,\ldots,A_k}_{m-i})$, where $A_k$ is any arbitrary agent with $A_k \neq A_i$.\edfn

Note that, if $u$ is additive for $A_i$, then we can simply denote $u$ with a vector $P=(p_1,\ldots,p_n)$ with $p_i\in\real^{\geq 0}$ representing the individual utilities added by receiving each resource. Formally, $u$ is unambiguously constructed from $P$ as follows: for all $\alpha=(\alpha_1,\ldots,\alpha_m)$, $u(\alpha)=\sum_{\{j\;|\;\alpha_j = A_i\}} p_j$. Thus, an additive utility function $u$ can be simply represented by that vector~$P=(p_1,\ldots,p_n)$. Given $r\in \real^{>0}$, we will say that an additive utility function $u$ is {\it $r$-limited} if $\sum_{1\leq i\leq n} p_i = r$ (in Section~\ref{sec:experiments}, experiments assuming and not assuming this constraint will be conducted).

Regarding non-additive utility functions, there are several ways to represent the preferences for {\it bundles} of resources:  extensionally defining the value of all possible {\it subsets} of resources, defining values to be added or subtracted if specific k-combinations of resources are owned, etc. This representation choice affects the complexity of any optimization problem based on them~\cite{nnr14}. Hereinafter, only additive utility functions will be considered in this paper.
%
%One possibility consists in extensionally defining the (positive or negative) utility {\it added} for all combinations of up to $k$ items, for some given constant $k\in\nat$. Similarly as the preferences for single items are given in the form of vectors of reals, the additional utilities due to pairs can be denoted by 2-dimension matrices, those due to 3-tuples by 3-dimension matrices, and so on up to $k$-dimension matrices. Let $P^1,\ldots,P^k$ be matrices over real numbers where each $P^h$ has $h$ dimensions, and let the value of $P^h$ at coordinates $j_1,\ldots,j_h$ be denoted $P^h_{j_1,\ldots,j_h}$. Then we can define $$u(\alpha)=\sum_{1\leq h\leq k} \sum_{\left\{j_1,\ldots,j_h \;\left| \footnotesize{\begin{array}{l} j_1 < \ldots < j_h,\\ \forall j_w : \alpha_{j_w} = A_i \end{array}}\right.\!\right\}} P^h_{j_1,\ldots,j_h}$$

%\noindent where index $h$ is used to range over all matrices $P^1,\ldots,P^k$, and indexes $j_1,\ldots,j_h$ range over all combinations of strictly increasing matrix coordinates denoting positions of resources actually assigned to agent $A_i$.\footnote{Indexes are required to be strictly increasing to avoid adding the same terms due to the corresponding combinations more than once. Thus, under this simple representation, many matrix cells of the full $h$-dimension matrix are irrelevant.}

%According to this representation of utility functions, a non-additive utility function $u$ including additional utilities added due to the combination of up to $k$ resources can be fully denoted by a tuple  $(P^1,\ldots,P^k)$. 

Let us denote by $\calU$ the set of possible additive utility functions. 
%of the type under consideration (i.e. additive, up to k-combinations, or any other kind). 
The utility functions of all agents in a tuple of agents $A$ will be denoted by a tuple $U=(u_1,\ldots,u_n)\in \calU^n$, where each $u_i$ is the utility function of agent $A_i$.

\newcommand{\eg}[2]{{\bf eg}_{#1}({#2})}

\newcommand{\egsol}[1]{{\bf egsol}({#1})}

We define the optimization problem of distributing resources in such a way that the utility of the agent receiving less utility is maximized. In the next definition, the auxiliary term $\eg{A,R,U}{\alpha}$ denotes the utility of the less satisfied agent when the distribution of resources is $\alpha$.

\bdfn\label{def:egal} Given $A=(A_1,\ldots,A_n)$, $R=(r_1,\ldots,r_m)$ and $U=(u_1,\ldots,u_n)$, for all $\alpha\in \calA$ let $\eg{A,R,U}{\alpha}=min\{u_j(\alpha)\;|\;1\leq j\leq n\}$. The egalitarian social welfare optimization problem consists in, given $A,R,U$, finding $\alpha$ maximizing $\eg{A,R,U}{\alpha}$. We will denote the solution of the problem for $A$, $R$, $U$ by $\egsol{A,R,U}$. 
\edfn

%In the rest of the paper, we will refer only to the particular case where utility functions are additive.
%
The NP-hardness of the previous problem is proved in~\cite{rr10}, and it cannot be approximated
in polynomial time within a factor of $\beta > 1/2$ unless $P = NP$~\cite{bd05}.
%
%\cite{lmm04}

\begin{proposition} Let us consider a variant of the egalitarian social welfare optimization problem where an additional input $r\in\real^{\geq 0}$ is given and all considered utility functions must be $r$-limited. The resulting problem is also NP-hard.\end{proposition} 

\begin{proof} In~\cite{rr10}, the NP-hardness of the problem in Definition~\ref{def:egal} is proved by a simple polynomial reduction from NP-hard problem PARTITION into it. In this reduction, the constructed instance of that problem contains only two agents having the same utility function, which in turn shows that the egalitarian optimization problem is NP-hard even under the additional constraint of taking only instances with two agents whose utility functions coincide. This implies the NP-hardness of a variant of the problem in Definition~\ref{def:egal} 
where only $r$-limited utility functions, for some $r\in\real$ given as problem input, can be used (i.e. when, for each agent, the addition of the valuations of all resources must be equal to a given constant $r$), as the aforementioned variant dealing with two agents having equal preferences (which is NP-hard) can be trivially polynomially reduced to that $r$-limited  problem: note that the addition of preferences of both agents are necessarily equal to some value $v$, so we can just set $r=v$. 
\end{proof}

%{\bf Inciso 1:} Conforme a~\cite{nnr14} (Tabla 1), se descubrio que el problema anterior es NP-completo con el modelo de preferencias no aditivas o aditivas (aka 1-aditivas) que estamos asumiendo en \cite{rr10} (tambien se menciona \cite{lmm04} pero no veo que tenga que ver), En el Teorema 5.1 de~\cite{rr10} se prueba con una reduccion muy sencilla desde PARTITION al caso 1-additive, o sea, simplemente aditivo. Respecto a la aproximabilidad, mirar la tesis doctoral~\cite{ngu13}: el caso aditivo puede verse en la seccion 3.2.2, y el caso k-aditivo parece ser la aportacion de la propia tesis. Si hablasemos del problema del pacto (no del trato) con preferencias aditivas, podriamos hacerlo asi. Asumamos que lograr una utilidad 0 para el participante mas perjudicado no es una solucion valida. Entonces podemos PTAS-reducir MAX ONES a dicho problema asumiendo que cada participante representa una clausula, y otro participante extra recibe una utilidad igual al numero de simbolos puestos a true. {\bf Hasta aqui el inciso.}

%{\bf Inciso 2:} La minimum-envy allocation mencionada en la pg 7 de~\cite{nnr14} no minimiza la diferencia de utilidad entre el mas beneficiado y el mas perjudicado, como creia, sino que minimiza la diferencia de utilidad entre recibir lo que me toca o recibir lo que le ha tocado a cualquier otro, que no es lo mismo (no miro lo feliz que es el otro, sino lo feliz que seria yo con lo suyo). Asi que nuestro concepto parece seguir siendo novedoso. Y mas si consideramos el pacto en vez del reparto, claro. {\bf Fin del inciso.}

\newcommand{\fake}[2]{{\bf fake}_{#1}({#2})}

\newcommand{\fakesol}[1]{{\bf fakesol}({#1})}

Let us call {\it auctioneer} to the third party responsible for, given the utility functions of all agents, finding the distribution of resources maximizing the egalitarian social welfare. Next we introduce the problem of finding the optimal fake utility function of a given agent, that is, the (potentially false) utility function letting that agent maximize its utility in an egalitarian social welfare distribution. That is, we search for the utility function such that, if agent $A_i$ sends it to auctioneer {\it and} the utility functions sent to the auctioneer by the other agents are those actually estimated by agent $A_i$, then the utility received by agent $A_i$ (with its {\it true} utility function) in the distribution maximizing the egalitarian social welfare is maximized. In the next definition, term $\fake{A,R,U,i}{f}$ denotes the utility achieved by agent $A_i$ if the utility function it sends to the auctioneer is $f$. Note that calculating this term implies solving the  optimization problem given in Definition~\ref{def:egal}, so {\it maximizing} that term means performing, in turn, an optimization of a term whose calculation requires another optimization.

\bdfn\label{def:fake} Given $A$, $R$, $U$ as before and $i\in\nat$, for all $f\in\calU$ let $\fake{A,R,U,i}{f} = u_i(\alpha_f)$ with $\alpha_f=\egsol{A,R,(u_1,\ldots,u_{i-1},f,u_{i+1},\ldots,u_n)}$. The optimal fake utility problem consists in, given $A$, $R$, $U$, and $i$, finding $f$ maximizing $\fake{A,R,U,i}{f}$. We will denote the solution of the problem for $A$, $R$, $U$, $i$ by $\fakesol{A,R,U,i}$.\edfn

Despite the fact that the previous problem intuitively consists in performing an optimization of a term whose calculation requires another optimization, we can see that solving it as it is defined is not difficult at all.

\begin{proposition}\label{prop:unlimited} Let the utility functions of each agent $A_j$ estimated by agent $A_i$ (and the utility of agent $A_i$ itself when $j=i$) be $P^j=(p^j_1,\ldots,p^j_n)$.
%, and let us suppose agent $A_i$ wants to find the best fake preferences to achieve the highest utility, assuming the other agents adhere to their real utility functions. 
The optimal fake utility function for $A_i$ is $P'^i=(c \cdot p^i_1,\ldots,c \cdot p^i_n)$ where $c\in\real^{>0}$ is any positive value such that $\sum_{1\leq s\leq n} c \cdot p^i_s < p^k_l$ for all $k\neq i$ and $l$ with $p^k_l > 0$.
\end{proposition}

\begin{proof}
First, let us suppose that there exists some allocation of resources $\alpha\in\calA$ such that $\eg{A,R,U}{\alpha} > 0$ (i.e. there is some allocation of resources giving a non-null utility to all agents). We show that the optimal fake utility function for $A_i$
is
%
%$P'^i=(c \cdot p^i_1,\ldots,c \cdot p^i_n)$ where $c\in\real$ is any positive value such that $\sum_{1\leq j\leq n} c \cdot p^i_j < p^k_l$ for all $k\neq i$ and $l$ with $p^k_l > 0$ 
%
the proposed one (note that $c$ always exists).
These fake preferences of agent $A_i$ are such that, even if the auctioneer gave all resources to $A_i$, then $A_i$ would still obtain less utility than the utility reached by any other agent by receiving any resource this agent has a  non-null preference for.
Note that, in this case, the goal of the auctioneer (i.e. maximizing $\eg{A,R,U}{\alpha}$) coincides with maximizing the {\it fake} utility of $A_i$ (i.e. preferences $P'^i$) {\it provided that} some non-null utility is given to the other agents (note that, if this constraint is impossible to achieve, then it contradicts our previous assumption that $\eg{A,R,U}{\alpha} > 0$ for some $\alpha$). On the other hand, note that the goal of $A_i$ is maximizing its {\it true} utility provided that {\it the same} constraint holds (as allocations not giving a non-null utility to all agents will never be picked by the auctioneer). Note that maximizing $P^i=(p^i_1,\ldots,p^i_n)$ subject to that constraint is equivalent to maximizing $P'^i=(c \cdot p^i_1,\ldots,c \cdot p^i_n)$ subject to the same constraint. Thus, the optimal strategy for $A_i$ consists in sending preferences $P'^i$ to the auctioneer.

In the (degenerate) case where there does not exist $\alpha$ with $\eg{A,R,U}{\alpha} > 0$, agent $A_i$ has no control over the allocation of resources picked by the auctioneer, because all of them are equally good according to the goal of the auctioneer (as any allocation $\alpha$ yields $\eg{A,R,U}{\alpha} = 0$). Thus, preferences $P'^i$ are as optimal for $A_i$ as any other preferences. \end{proof}

Let us consider the problem given in Definition~\ref{def:fake} under the additional constraint that  all agent preferences are required to be $r$-limited for some $r\in\real^{>0}$ given as problem input. In this case solutions cannot be trivially constructed as before (note that $P'^i$ would not respect the constraint). Actually it seems that, in this case, the optimization of a term whose calculation requires another optimization {\it does} make the resulting problem much harder. We conjecture the resulting (decision) problem is $\Sigma_2^P$-complete. 

\section{Experiments}\label{sec:experiments}
In this section we describe several experiments we have conducted to check the usefulness of lying in an egalitarian social welfare scenario. That is, we study what happens when an agent sends to the auctioneer a preference vector $P$ different to its real preferences. From now on, we will denote by $T$ (True) the vector of real preferences of a given agent for the available resources. On the other hand, we will denote by $M$ (Modified) the modified preference vector, i.e. the one actually communicated to the auctioneer.

We will consider three stages in our experiments. First, we will analyze the usefulness of some predefined lying strategies (e.g. increase/decrease our preferences about the resources we value the most or the resources valued the most by other users, etc.); second, we will use genetic algorithms to dynamically search for the optimal false preferences for each specific problem instance (instead of composing them according to predefined rules as before); and third, we will analyse the performance of our lies when the information we have about the preferences of other users is not perfect.

According to the restrictions the agents have on the preferences they are allowed to communicate, for each of the aforementioned stages we consider two scenarios. In the first one, denoted ``unlimited scenario,'' each agent must establish a preference for each resource so that this value is within the interval $(0, 100)$. In the second case, denoted as ``limited scenario,'' in addition to the previous restriction, the sum of all the preferences is required to be equal to $100$.

In all experiments, it is assumed that Agent $1$ is the lying agent we are analysing. In each case, this agent will try to achieve the greatest possible utility by modifying its communicated preferences. We performed different sets of experiments with different numbers of agents and available resources, and we observed the same trends and patterns in all of them. Hence, for the sake of conciseness and specificity in the data exposition, we will only show the results for the case study where ten resources are to be distributed among four agents. Table~\ref{Tab:Table1} contains the real preferences of the four agents in the unlimited scenario, whereas Table~\ref{Tab:Table2} contains the preferences of the agents in the limited case. As we said before, our experiments will analyse what happens when Agent $1$ communicates different preferences. However, in any case, the usefulness obtained by Agent $1$ will be computed by using its actual preferences, that is, those shown in the corresponding row of Table~\ref{Tab:Table1} and Table~\ref{Tab:Table2}.

% Table 1
\renewcommand{\arraystretch}{1.5}
\setlength{\tabcolsep}{3pt}
\begin{table}[t]
\begin{center}
    \caption{Real preferences of the agents in the unlimited scenario}
\scriptsize
%\footnotesize
    \begin{tabular}{|c|cccccccccc|}
        \hline
 %       \rowcolor[HTML]{EFEFEF} 
 %       \multicolumn{11}{|c|}{\cellcolor[HTML]{EFEFEF}\textbf{Preferences in unlimited scenario}}  \\ \hline
         & 1     & 2     & 3     & 4     & 5     & 6     & 7     & 8     & 9     & 10    \\ \hline
        \rowcolor[HTML]{E7F0FB} 
        Agent 1  & 42.36 & 19.18 & 75.37 & 42.32 & 60.20 & 68.98 & 85.30 & 20.66 & 31.67 & 60.41 \\
        Agent 2  & 53.70 & 96.73 & 70.87 & 68.47 & 86.95 & 51.34 & 3.06  & 30.01 & 55.03 & 45.42 \\
        Agent 3  & 88.49 & 17.88 & 70.73 & 98.71 & 30.42 & 40.71 & 98.41 & 85.19 & 42.08 & 57.05 \\
        Agent 4  & 95.08 & 50.57 & 69.31 & 26.01 & 56.03 & 64.09 & 55.08 & 6.62  & 49.24 & 31.69  \\ \hline
    \end{tabular}
\label{Tab:Table1}
\end{center}    
\end{table}

% Table 2
\renewcommand{\arraystretch}{1.5}
\begin{table}[t]
\begin{center}
    \caption{Real preferences of the agents in the limited scenario}
\scriptsize
%\footnotesize
    \begin{tabular}{|c|cccccccccc|}
        \hline
%        \rowcolor[HTML]{EFEFEF} 
%        \multicolumn{11}{|c|}{\cellcolor[HTML]{EFEFEF}\textbf{Preferences in limited scenario}}    \\ \hline
%        Resource 
        & 1     & 2     & 3     & 4     & 5     & 6     & 7     & 8     & 9     & 10  \\ \hline
        \rowcolor[HTML]{E7F0FB} 
        Agent 1  & 17.67 & 12.58 & 4.35  & 12.00 & 5.47  & 0.77  & 14.57 & 3.49 & 16.55 & 12.504 \\
        Agent 2  & 1.927 & 6.09  & 19.66 & 18.17 & 10.51 & 10.75 & 2.42  & 3.31 & 23.04 & 4.07   \\
        Agent 3  & 16.95 & 12.24 & 11.64 & 7.33  & 7.63  & 12.57 & 9.87  & 8.73 & 11.33 & 1.66   \\
        Agent 4  & 14.41 & 2.20  & 16.78 & 1.03  & 13.15 & 9.70  & 5.18  & 5.46 & 17.19 & 14.85 \\ \hline
    \end{tabular}
\label{Tab:Table2}
\end{center}    
\end{table}

\subsection{Testing generic lying strategies}
In order to achieve a greater benefit, Agent $1$ may lie when it communicates its preferences to the auctioneer. Note that, when Agent $1$ communicates its lie, in general it does not certainly know if that lie will provide it with a greater benefit, since at that moment the distribution of resources has not been made yet. Next we investigate if Agent $1$ has some kind of lying strategy to follow that will generally provide it with profitable lies. 

The aim of our first set of experiments is to test a set of predefined lying strategies. For instance, is it a good strategy to always communicate lower (or greater) preferences for the item we are most interested in? Is it useful to follow the same strategy for the item we are less interested in (or for the items supposed to be less valued by other agents)? 

We have tested different strategies. For each of them, we compare the results obtained when the agent does not lie and when the agent uses such strategy to lie up to different {\it lying levels} (higher levels mean higher deviations from the corresponding true values). Table~\ref{Tab:TableStrategies} shows some examples of strategies that have been applied.

% Table 3
\renewcommand{\arraystretch}{1.5}
\setlength{\tabcolsep}{1.0pt}
\begin{table}[t]
   \caption{Summary of basic lying strategies}
    \footnotesize
    \begin{tabular}{cl}
    \rowcolor[HTML]{EFEFEF} 
\   {Test} & \multicolumn{1}{c}{\cellcolor[HTML]{EFEFEF}\textbf{Description}}                  \\ \hline
1             & Decrease the preference for random resources                             \\
2             & Increase the preference for random resources                              \\
3             & Increase preference for the resources most valued by other agents           \\
4             & Increase preference for the resources less valued by other agents \\
5             & Increase preference for the resources most valued by itself             \\
6             & Increase preference for the resources less valued by itself             \\
7             & Decrease preference for the resources most valued by other agents           \\
8             & Decrease preference for the resources less valued by other agents \\
9            & Decrease preference for the resources most valued by itself             \\
10            & Decrease preference for the resources less valued by itself           
    \end{tabular}
 \label{Tab:TableStrategies}
\end{table}

%To this purpose, we have designed some experiments in which Agent 1 modifies its preference vector before communicating it to the deliverer. As we know, Agent 1 can modify its preferences as long as it does not violate the constraints of the way we are working.

In order to calculate the benefit (or loss) Agent~$1$ would obtain by communicating its lie, first we must know the utility it obtains when it communicates its real preferences. Since obtaining the optimal resource allocation according to the egalitarian social welfare (see Definition~\ref{def:egal}) is an NP-hard problem, we use a genetic algorithm to deal with this problem. The algorithm in charge of finding the solution to this problem 
%was defined as $\egsol{A,R,U}$, where in this case $U$ corresponds to the real preference vector $T$ that Agent $1$ has, 
will be called LLGA ({\it Lower-Level Genetic Algorithm}). 
In our examples, the population of LLGA will have 50 individuals, running 50 iterations and using tournament selection.

On the other hand, we also have to solve the same problem when actually using the corresponding lie, which is given by the preference vector $M$. Note that doing so will potentially provide a different solution. To know the benefit or loss that Agent $1$ obtains by using its lie, it is enough to subtract the utility it obtains with the second solution (i.e. when it is using its vector $M$) minus the utility it obtains with the initial solution (i.e. when it is using its vector $T$). If the result of such subtraction is positive, then it is worth communicating the lie to the auctioneer. 

Let us start with the most simple scenario, that is, the unlimited case. In the following figures, the y-axis represents the profit (or loss) obtained by Agent $1$, whereas the x-axis represents the degree of application of the strategy (i.e. the amount of increase/decrease in which the communicated preferences differ from the preferences it actually has for the resources). These degrees of application are represented by 100 progressive increases/decreases.

\begin{figure}[t]
\includegraphics[width=9cm]{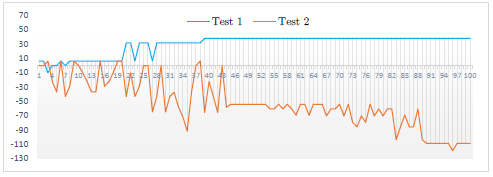} 
\caption{Results obtained with generic lying strategies in the unlimited scenario: decreasing (test 1) or increasing (test2) preferences}\label{Fig:figureLLGAUnlimited1}
\end{figure}

Figure~\ref{Fig:figureLLGAUnlimited1} shows the results of applying our first two %four 
tests, that is,  decreasing and increasing, respectively, the preferences concerning some random resources. %first eight resources. %The utility obtained by Agent $1$ when using its real preference vector $T$ was $214.71$. 
As it can be seen, when we start modifying a little bit the preferences %for the first five resources 
of some of the resources, %the difference is not remarkable. None 
none of the strategies (decreasing or increasing preferences) seems to work. However, increasing preferences is clearly worse, as some negative peaks start appearing near from the beginning.
% during the first modifications. %, although decreasing preferences leads to considerably greater falls. 
These peaks in the graph are due to the amount of resources we are working with. Let us recall that we only have ten resources and that varying just one of them in a  solution may make results suffer great variations (if more resources were considered, then these peaks would soften and we would see much more progressive variations). At the same time, as Agent $1$ keeps modifying its preferences more, we observe how the strategy followed in Test $2$ implies greater losses of utility. However, during Test $1$ the utility tends to stabilize. These strategies suggest that communicating a lie with higher preferences with respect to the preference vector $T$ seems to be a bad strategy for Agent $1$ to follow. On the other hand, as Test $1$ lies %about a greater amount 
more about the %of resources by communicating lower 
preferences with respect to the preference vector $T$, the utility obtained is considerably increased, which provides a benefit to Agent~$1$.

\begin{figure}[t]
\includegraphics[width=9cm]{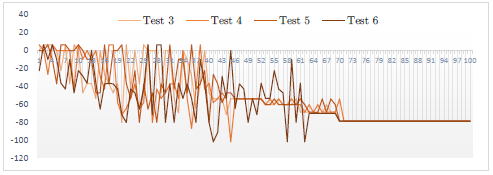} 
\caption{Results obtained with generic lying strategies in the unlimited scenario: different strategies increasing preferences}\label{Fig:figureLLGAUnlimited2}
\end{figure}

Let us look at some more tests where the preference for resources is increased. Figure~\ref{Fig:figureLLGAUnlimited2} shows that these tests do not provide any benefit to Agent $1$. It seems that none of the lies in which higher preferences are communicated to the auctioneer provide benefit, which reinforces the results of Test $2$ that we saw earlier.

\begin{figure}[t]
\includegraphics[width=9cm]{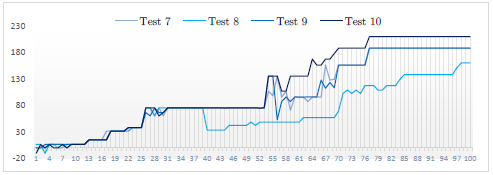} 
\caption{Results obtained with generic lying strategies in the unlimited scenario: different strategies decreasing preferences}\label{Fig:figureLLGAUnlimited3}
\end{figure}

On the other hand, Figure~\ref{Fig:figureLLGAUnlimited3} shows some strategies in which Agent $1$ communicates preferences that are lower than its real preferences. As we intuited, all strategies that involve decreasing preferences work, and they also become better as we consider a greater amount of resources. As we see, some of these strategies are better than others, especially Test $10$, where Agent $1$ decreases preferences for the %eight 
resources it values less.

%{\bf (el parrafo que viene a continuacion y la figura igual la quereis quitar. Lo he hecho tan solo para ver que si seguimos disminuyendo cada vez las preferencias al final llegaremos a la utilidad optima, aunque por otro lado en este punto del articulo aun no se sabe que es eso de la ultilidad optima...)}

These last strategies make us wonder if the utility achieved during Test $10$ is the highest possible. Despite being very close, it is not the maximum. By decreasing the preferences for all resources, a slightly better result is obtained. %However, it is necessary to continue decreasing the preferences during some more iterations. %Figure~\ref{Fig:figureLLGAUnlimited4} shows it is possible to achieve even greater utility when Agent 1 decreases all its preferences. 
As we will see later, this last test verifies the best strategy that Agent $1$  has to lie (accordingly to Proposition~\ref{prop:unlimited}).

%\begin{figure}[t]
%\includegraphics[width=9cm]{images/figureLLGAUnlimited4.PNG} 
%\caption{Results obtained with generic lying strategies in the unlimited scenario: Decreasing the preferences about all the resources}\label{Fig:figureLLGAUnlimited4}
%\end{figure}

\begin{figure}[t]
\includegraphics[width=9cm]{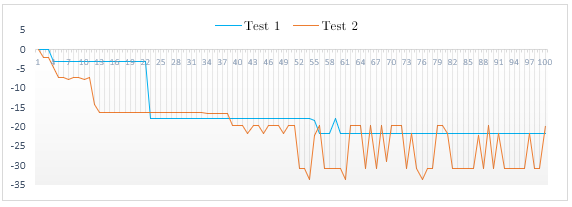} 
\caption{Results obtained with generic lying strategies in the limited scenario: decreasing (test 1) or increasing (test2) preferences}\label{Fig:figureLLGALimited1}
\end{figure}

\begin{figure}[t]
\includegraphics[width=9cm]{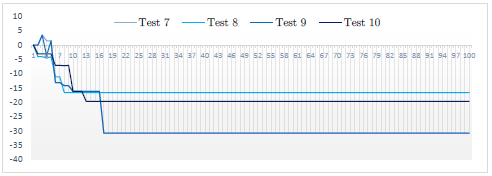} 
\caption{Results obtained with generic lying strategies in the limited scenario: different strategies decreasing preferences}\label{Fig:figureLLGALimited3}
\end{figure}

All in all, these tests clearly indicate that it is trivial to lie in the unlimited scenario, because a good strategy to follow by Agent $1$ is to decrease its preferences for as many resources as possible. However, when we deal with the {\it limited} scenario, it is not possible to decrease the preferences concerning all the resources. Thus, each time a preference is lowered, it is necessary to increase the preference about another resource, so that the total addition keeps constant. Figure~\ref{Fig:figureLLGALimited1} shows the results obtained with the same strategies as before, but in the case of the limited scenario. 
%\textbf{(se que parece mucha casualidad la Figura 5, pero ha salido asi de parecida la grafica cuando se aumentan y disminuye las preferencias los por ocho primeros recursos. Aunque en realidad tiene sentido porque si hacemos lo mismo (en el caso limitado) no es de extraniar que la utilidad varie de la misma forma. Aunque haya sido asi, si esto queda muy raro puedo ejecutar de nuevo las pruebas y tal vez el resultado sea algo distinto)}  %In this case, the utility obtained by Agent $1$ when using its real preference vector $T$ was $34.23$.
If we look at it, we realize that the new constraint makes it very difficult to find a strategy that brings benefits to Agent $1$. In fact, all of them have turned out to be unsuccessful strategies, producing a reduction in the satisfaction of Agent $1$. 
Even when Agent $1$ decreases the preferences for the resources it values the most (see Figure~\ref{Fig:figureLLGALimited3}), it has to increase its preferences for other resources. Thus, the auctioneer can assign Agent $1$ some resources which are not really very useful for that agent.

\begin{figure}[t]
\includegraphics[width=9cm]{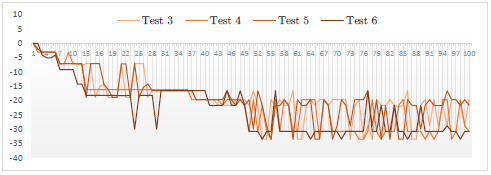} 
\caption{Results obtained with generic lying strategies in the limited scenario: different strategies increasing preferences}\label{Fig:figureLLGALimited2}
\end{figure}

In other words: there does not seem to be a generic strategy to be followed by Agent $1$ in the limited scenario so that it consistently improves its outcome when it lies. However, this does not necessarily imply that lying is completely discouraged ---it only means that generic, predefined strategies do not seem to be useful. However, for each {\it specific} case study it could be possible to dynamically find a tailored lie that provides a higher utility to Agent $1$.
In fact, in our next experiments we will analyse such a possibility.

%Considering these tests (and many others) it is considerable to think that the Limited mode manages to completely discourage lying, since by using its preference vector R, Agent 1 achieves a utility of 48.8. In other words: there does not seem to be a strategy to follow by Agent 1 so that when it comes to lying it achieves benefits. Have we really managed to completely disincentive the lie? Is there no lie with which to obtain greater utility than when Agent 1 simply tells the truth? {\bf (estas dos preguntas me suenan muy mal, pero no se como ponerlas)} To answer these questions we must take a more ambitious step in our experiments.

\subsection{Looking for the best lie}
We have seen that generic lying strategies are useful in the unlimited scenario (in fact, reducing the valuation of all resources is a good strategy in any unlimited scenario), but we have not been able to find a good strategy in the case of the limited scenario.
%
%We have seen some of the strategies that Agent 1 can follow when he communicates a lie to the auctioneer to achieve greater utility than he would get by communicating his actual preferences. A priori {\bf (esto se dice igual?)}, it seems that such strategies only bring benefits in the Not limited mode and thanks to the restriction imposed {\bf (queda bien poner imposed?)} in the Limited mode, we have managed to completely disincentive the lie.

Next we will try to construct {\it ad hoc} lies to deal with each specific configuration. In fact, now our aim is to find the {\it optimal} lie for each case study. That is, we want to find the best lie Agent $1$ can communicate, i.e. the lie with which it would achieve the greatest possible utility. Obtaining this utility in the unlimited scenario is trivial: regardless of the number of agents and resources we have, Agent $1$ achieves the highest utility when it proportionally underestimates the valuation of its own preferences so much as to make the auctioneer think Agent $1$ will be the least satisfied agent in any distribution giving a non-null utility to all agents (see Proposition~\ref{prop:unlimited}). In this case, Agent $1$ will be assigned the {\it n}-{\it m} resources that it values the most with its real preferences, where {\it n} is the number of available resources and {\it m} is the number of other agents. Note that giving some satisfaction to the other agents requires giving at least one resource to each agent. However, in order to maximize the utility received by Agent $1$, the rest of the agents must receive those resources Agent $1$ values the least. By taking to the extreme the strategy of decreasing its preferences when lying, Agent $1$ will achieve the remaining resources, which are those it values the most. 
%During this second part of the experiments we are going to pursue {\bf (pursue lo he tenido que traducir y no se si queda bien)}  a more ambitious objective: to find the best lie that Agent 1 can communicate, that is to say, the lie with which he would achieve the greatest possible utility. Calculating this utility in the Not limited mode is trivial: regardless of the number of agents and resources we have, the greatest utility that Agent 1 can achieve is to lie in such a way that he is assigned the {\it n}-{\it m} resources that he most values (with his real preferences), where {\it n} is the number of resources available and {\it m} the number of agents remaining. Note that, for the other agents to be "somewhat"  {\bf (dejo esta palabra o podria poner "rather"?)}  satisfied, they should receive at least one resource. However, it is not useful to assign them just any resource. To maximize the utility that Agent 1 receives, the rest of the agents should receive those resources that Agent 1 values less. In this way and taking to the extreme his strategy of decreasing his preferences when lying, he could achieve that the rest of the resources (which are precisely the ones that he values the most) are assigned to him.

In order to find the best lie, we must solve the problem of finding the utility function $f$ maximizing $\fake{A,R,U,i}{f}$, where $f$ denotes the false preferences to be sent to the auctioneer (see Definition~\ref{def:fake}). The resolution of this problem, as we already saw, involves making an optimization of an optimization. The genetic algorithm in charge of finding this best lie is called ULGA ({\it Upper-Level Genetic Algorithm}). In this algorithm, each individual of the population represents a possible lie of Agent $1$. That is, an individual is a vector $M$ representing a possible valuation of resources Agent $1$ can communicate to the auctioneer. The fitness function used in ULGA to evaluate the quality of individuals (i.e. of candidate false preferences) requires evaluating the usefulness of the corresponding lie $M$ to provide profit to Agent~$1$. Thus, for each individual $M$, computing its fitness firstly requires computing the optimal distribution of resources the auctioneer can perform according to the rules of egalitarian social welfare. Consequently, for each lie $M$ in the population of ULGA, we use algorithm LLGA to compute the  distribution of resources picked by the auctioneer when the preferences sent by Agent~$1$ are those given in $M$. The fitness of each possible individual (lie) $M$ in ULGA is computed by adding the {\it real} values Agent~$1$ would get for each resource assigned to Agent~$1$ in the distribution of resources found by LLGA (note that those {\it real} values are those in vector $T$, not in $M$).

%In order to 
To check that our method obtains good solutions, we start with %by considering 
the most simple case, the unlimited scenario. Then, we will move to the most interesting case, the limited scenario.

%Figure~\ref{Fig:figureULGAUnlimited} shows a graph where the y-axis indicates the achieved utility  (showing the benefit/loss over the utility obtained by Agent $1$ when using its true preference vector $T$) and the x-axis shows the number of times the population of ULGA evolves.

%\begin{figure}[t]
%\includegraphics[width=9cm]{images/figureULGAUnlimited.PNG} 
%\caption{ULGA obtaining the best lie in unlimited scenario}\label{Fig:figureULGAUnlimited}
%\end{figure}

The population of 50 individuals was evolved 3,000 times
using tournament selection.
% although, as we see, the maximum utility was reached in evolution step 2,835. 
Both the optimal lie (preferences) for Agent 1 and the reached distributions  can be checked in Table~\ref{Tab:Table3} (in the distributions, numbers denote the agent each resource is assigned to).
As it can be seen, all resources but three are assigned to Agent~$1$. Moreover, those three resources are the ones Agent $1$ likes the least. Hence, ULGA  actually found the solution providing the best distribution of resources for Agent $1$, according to the obvious theoretical limit that at least one resource has to be assigned to each agent. Let us remark that ULGA does not need to decrease {\it forever} the preferences of Agent~$1$ for the resources in order to achieve continuously better lies. Actually, the best possible distribution of resources for Agent~$1$ is reached when its valuations are  {\it low enough} (note that this experimental observation is consistent with Proposition~\ref{prop:unlimited}).

%As we said before, the utility perceived by Agent 1 must be calculated based on his real preferences and not on his lie. With these results, we can affirm that we have reached the theoretical optimum in this test, which is 434,9889 (remember that the graph shows the benefit/loss, not the total utility that Agent 1 perceives).

% Table 
\renewcommand{\arraystretch}{1.1} %{1.5}
\setlength{\tabcolsep}{3pt}
\begin{table}[t]
\begin{center}
\caption{Best lie obtained by ULGA in the unlimited scenario}
\scriptsize
%\footnotesize
    \begin{tabular}{|
        >{\columncolor[HTML]{EFEFEF}}c |cccccccccc|} \hline
        %Resource 
        & 1     & 2     & 3     & 4     & 5     & 6     & 7     & 8     & 9     & 10  \\ \hline
        %Real preferences of Agent $1$  & 42.36 & 19.18 & 75.37 & 42.32 & 60.20 & 68.98 & 85.30 & 20.66 & 31.67 & 60.41 \\
        Real A$1$  & 42.36 & 19.18 & 75.37 & 42.32 & 60.20 & 68.98 & 85.30 & 20.66 & 31.67 & 60.41 \\
        Distribution & 4    & 2    &1     & 3    & 2    &4    & 1     & 3    & 4   & 1   \\ \hline
        Lie A$1$    & 2.21 & 2.27 & 10.57 & 3.12 & 6.5 & 5.55 & 12.09 & 2.18 & 2.27 & 4.87 \\
        Distribution & 1    & 4    & 1     & 1    & 1    & 1    & 1     & 3    & 2    & 1   \\ \hline
    \end{tabular}
    \label{Tab:Table3}
\end{center}    
\end{table}

Regarding the limited scenario, the population evolved the same number of times, but in this case %we observed that our generic strategies were not useful. Consequently, finding useful lies by using ULGA was  expected to be harder. 
%Figure~\ref{Fig:figureULGALimited} shows the evolution of the quality of the lies found by ULGA.
%It can be seen that 
the utility benefit is relatively lower, obtaining only a profit of a little bit more than 10. This is something we already suspected would happen, since none of the basic predefined strategies worked in our first experiments, as we saw in the previous section.
Anyway, the experiments prove it is still profitable to lie in the limited scenario, although it is clearly harder to find a useful lie to be communicated to the auctioneer. 
Table~\ref{Tab:Table4} shows the preference vector $T$ of Agent $1$ and the best lie found by the ULGA, as well as the corresponding distributions for each kind of preference. Let us remark that the best lie obtained is very different with respect to the real preferences. Moreover, there is not a clear general strategy to move from the real preferences to the best lie, as it depends on small adjustments related with the preferences of the rest of agents.

%\begin{figure}[t]
%\includegraphics[width=9cm]{images/figureULGALimited.PNG} 
%\caption{ULGA obtaining the best lie in the limited scenario}\label{Fig:figureULGALimited}
%\end{figure}

% Table 4
\renewcommand{\arraystretch}{1.1} %{1.5}
\setlength{\tabcolsep}{3pt}
\begin{table}[t]
\begin{center}
\caption{Best lie obtained by ULGA in the limited scenario}
\scriptsize
%\footnotesize
    \begin{tabular}{|
        >{\columncolor[HTML]{EFEFEF}}c |cccccccccc|}  \hline
        %Resource 
        & 1     & 2     & 3     & 4     & 5     & 6     & 7     & 8     & 9     & 10  \\ \hline
        %Real preferences of Agent $1$ 
        Real A$1$ & 17.67 & 12.58 & 4.35  & 12.00 & 5.47  & 0.77  & 14.57 & 3.49 & 16.55 & 12.504 \\
        Distribution & 1    & 3    & 2     & 2    & 4    & 3    & 4     & 3    & 1    & 4    \\ \hline
        %Best lie of Agent $1$    
        Lie A$1$ & 13.53 & 9.52 & 14.35 & 9.86 & 6.1 & 8.83 & 9.03 & 5.43 & 12.2 & 11.11 \\
        Distribution & 1    & 3    & 2    & 2   & 4    & 3    & 1     & 3    & 1    & 4    \\ \hline
    \end{tabular}
    \label{Tab:Table4}
\end{center}    
\end{table}

\subsection{Analysing imprecise information}
The previous experiments have shown that, even in the limited scenario, it is possible to find ad hoc lies making Agent~$1$ reach a higher profit if communicated to the auctioneer. However, the ad hoc solutions found by ULGA require that Agent $1$ knows the preferences of the rest of agents, as they are needed to evaluate the usefulness of each lie when applying LLGA. Hence, a new question arises: is it still a good idea to use the lie obtained by ULGA when we are not completely sure as to whether the preferences of the rest of agents are those actually assumed by Agent~$1$ and introduced in the algorithm? That is, what happens when our predictions about the rest of agents are not good enough?

In this section we report some experiments conducted to analyze the aforementioned problem. We used ULGA to find the optimal lie Agent~$1$ has to communicate to the auctioneer, assuming the preferences of the other agents are those considered by Agent~$1$. Then, we evaluated the usefulness of this lie when such preferences are not correct. We evaluate different situations depending on the accuracy of the  prediction of the preferences of the other agents assumed by Agent $1$. In particular, we randomly generated the {\it real} preferences of the other agents from the preferences  assumed by Agent~$1$ by applying a normal distribution whose average value is at the assumed preference. We generated several new instances where the standard deviation of such distribution is small, next other instances where it was slightly bigger, and so on for a variety of standard deviations.

\begin{figure}[t]
\includegraphics[width=9cm]{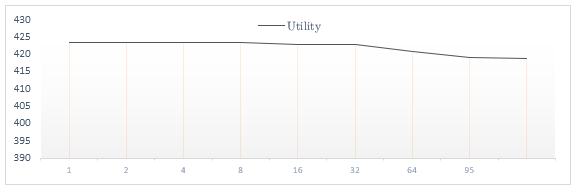} 
\caption{Lying in the unlimited scenario with imprecise information}\label{Fig:figureNormalUnlimited}
\end{figure}

\begin{figure}[t]
\begin{center}
\includegraphics[width=9cm]{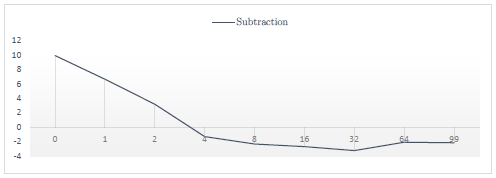} 
\end{center}
\caption{Lying in the limited scenario with imprecise information}\label{Fig:figureNormalLimited}
\end{figure}

In order to have a better view of how the utility of Agent~$1$ is affected by the imprecision, Figures~\ref{Fig:figureNormalUnlimited} and~\ref{Fig:figureNormalLimited} show the utility improvement obtained when using the lie found by ULGA in the context of imprecise information. The $x$-axis represents the standard deviation of the normal distributions used to randomly construct the actual preferences of the other agents from the corresponding assumed preferences.

%In addition, to know what happens to that utility, as we apply a larger deviation it is necessary to calculate a new utility. 

For each new deviation, 1,000 alternative utility scenarios are created by using the corresponding deviation. For each of them, we computed the benefit of using the real preferences of Agent~$1$ (i.e. not lying) and the benefit Agent~$1$ obtains by using the lie provided by ULGA when it uses the preferences of the rest of agents assumed by Agent~$1$ (that is, the preferences before applying the normal distribution). Then, we compute the average behaviour of these 1,000 cases.

Again, let us firstly discuss the unlimited scenario. Figure~\ref{Fig:figureNormalUnlimited} shows the obtained results. Specifically, the graph shows how the utility obtained by Agent $1$ evolves when it uses its lie and, at the same time, we are increasing the deviation yielding the actual preferences of the other agents. As  it  can  be  seen,  the  line  stays relatively  horizontal  as  we  modify  the  deviation  until $\sigma$ = 32,  which  implies  that  the  deviation  produces  the  same effect when applied on the different preferences. From this point  on,  the  utility is slightly decreased.  This  means  that  Agent  $1$'s  lie  continues  to  be efficient (at least, more efficient than telling the truth) even when the preferences of the other agents differ from those it estimated. Note that, even when we apply a normal distribution with $\sigma$ = 99.0, the utility is still clearly above the utility obtained with the preference vector $T$ of Agent $1$ (which is 221.08 and can be calculated by using Table~\ref{Tab:Table3}).  This result shows that, independently of the  estimation  made by  Agent $1$ about the  preferences of the other agents, it  is always  worth lying to the  auctioneer. %  by  communicating  the  lowest  possible preferences for the available resources.

If we move to the limited scenario, in Figure~\ref{Fig:figureNormalLimited} we see very different results from those obtained in the unlimited scenario. 
This time, let us denote by A the total utility achieved by Agent~$1$ when using  the lie  provided  by  ULGA. In addition, let us denote by A' the total utility achieved by Agent 1 when it uses its real preference vector T. If we calculate the result of A'-A, we will see that after a certain deviation it is not worth for Agent 1 to communicate its lie to the auctioneer. Initially, the result of the subtraction keeps positive for small deviations. However, even with a relatively small deviation ($\sigma = 4$), the profit obtained by Agent 1 when lying is worse than when saying the truth.
This result shows that it is worth lying when the preferences of the other agents are very similar to those Agent~$1$ estimated but, otherwise, there is not incentive to try to lie. 
That is, according to the experimental results, in the limited scenario it is worth trying to lie only provided that the estimations about the preferences of the rest of agents are very close to their actual preferences.

\section{Conclusions and future work}\label{sec:conclusions}

The experimental results we have obtained in the unlimited scenario let us conclude that the best strategy to be used by Agent $1$ is to lie by communicating a high underestimation of its preferences for the available resources (which is consistent with Proposition~\ref{prop:unlimited}). This way, the auctioneer will be forced to assign the {\it n}-{\it m} resources that it values the most to Agent~$1$, so that Agent $1$ will achieve the greatest possible utility. On the other hand, the tests carried out in the limited scenario allow us to conclude that, in this case, none of the considered predefined strategies  works. It seems that there is no strategy to be followed by Agent $1$ other than simply testing many combinations by decreasing and increasing its preferences. In spite of this, the limited scenario does not succeed in completely eliminating the lie, because our ULGA algorithm manages to find fake preferences for Agent $1$ providing a higher profit than sending the true preferences. Also, lying in the unlimited scenario remains the best option even when the estimation made by Agent $1$ about the preferences of its mates differ greatly from the preferences these agents actually have. In other words: in the unlimited scenario, it is always worth lying. On the contrary, in the limited case there seems to be no clear strategy to lie. In fact, we can only find good lies to be communicated to the auctioneer provided that our estimations about the preferences of the rest of participants are really good. Otherwise, relatively small deviations from our estimations make our lies obtain results being worse than those achieved when revealing our real preferences.

Summarizing, although lying is not completely disincentivized in our limited scenario, we can conclude that, in practical terms, lying is not useful in such scenario, as we would need very detailed information about the preferences of the rest of participants. Thus, a simple rule modification consisting in forcing the addition of all preferences be equal to some constant can be very useful in practical terms to avoid lying in the egalitarian social welfare.

%Regarding our lines of work, we are currently 
%studying the computational complexity of the problem of finding the optimal lie in the limited scenario (as pointed out before, we conjecture it will be $\Sigma_2^P$-complete). We are also studying the situation where non-additive utilities can be defined by using packages of resources. 

\comen{
\textbf{Quitar el resto desde aqui}
Some lines for our future work focus on making small variations on the rules of our delivery while others continue to explore scenarios seen with experiments that go one step further. Some of these lines are as follows:

\begin{itemize}
%\item Study what happens when both number of agents and resources available are changed.
\item Study how the distribution changes when considering non-additive utilities defined by packages of resources. In this case, the utility obtained by taking certain resources is not only the sum of the preferences that there were towards them, but there is an added (or reduced) utility according to which combinations of resources the agents take.
\item  Study if there is a method that completely discourages lying in the limited scenario.
\item  In relation to the previous point, study what happens when several agents try to use their respective best lies at the same time.
\item Study the computational complexity of finding the optimal lie in the limited scenario.
\end{itemize}

 }

\bibliographystyle{IEEEtran}

\bibliography{tse_bib}

@article{xing2014innovative,
  title={Innovative computational intelligence: a rough guide to 134 clever algorithms},
  author={Xing, Bo and Gao, Wen-Jing},
  year={2014},
  publisher={Springer}
}

@article{RR19water,
  title={Water-based metaheuristics: How water dynamics can help us to solve NP-hard problems},
  author={Rubio, Fernando and Rodr{\'\i}guez, Ismael},
  journal={Complexity},
  volume={2019},
  publisher={Hindawi}
}

@article{RRR17JoCS,
  title={Applications of river formation dynamics},
  author={Rabanal, Pablo and Rodr{\'\i}guez, Ismael and Rubio, Fernando},
  journal={Journal of computational science},
  volume={22},
  pages={26--35},
  year={2017},
  publisher={Elsevier}
}

@book{chen2018oxford,
  title={The Oxford handbook of computational economics and finance},
  author={Chen, Shu-Heng and Kaboudan, Mak and Du, Ye-Rong},
  year={2018},
  publisher={Oxford University Press}
}

@inproceedings{marketing16,
  title={Automatic media planning: optimal advertisement placement problems},
  author={Rodr{\'\i}guez, Ismael and Rubio, Fernando and Rabanal, Pablo},
  booktitle={2016 IEEE Congress on Evolutionary Computation (CEC)},
  pages={5170--5177},
  year={2016},
  organization={IEEE}
}

@inproceedings{monaco19,
  title={On the performance of stable outcomes in modified fractional hedonic games with egalitarian social welfare},
  author={Monaco, Gianpiero and Moscardelli, Luca and Velaj, Yllka},
  booktitle={AAMAS'19},
  pages={873--881},
  year={2019}
}

@article{grid05,
  title={The grid economy},
  author={Buyya, Rajkumar and Abramson, David and Venugopal, Srikumar},
  journal={Proceedings of the IEEE},
  volume={93},
  number={3},
  pages={698--714},
  year={2005},
  publisher={IEEE}
}

@article{leon10,
  title={Using economic regulation to prevent resource congestion in large-scale shared infrastructures},
  author={Le{\'o}n, Xavier and Trinh, Tuan Anh and Navarro, Leandro},
  journal={Future Generation Computer Systems},
  volume={26},
  number={4},
  pages={599--607},
  year={2010},
  publisher={Elsevier}
}

@article{winnerDet06,
  title={The winner determination problem},
  author={Lehmann, Daniel and M{\"u}ller, Rudolf and Sandholm, Tuomas},
  journal={Combinatorial auctions},
  pages={297--318},
  year={2006}
}

@article{VickreyGeneral99,
  title={A generalized {V}ickrey auction},
  author={Ausubel, Lawrence M.},
  journal={Univ. Maryland},
  year={1999}
}

@inproceedings{VickreyMalo96,
  title={Limitations of the Vickrey auction in computational multiagent systems},
  author={Sandholm, Tuomas W},
  booktitle={ICMAS-96},
  pages={299--306},
  year={1996}
}

@article{VickreyMalo07,
  title={13 reasons why the Vickrey-Clarke-Groves process is not practical},
  author={Rothkopf, Michael H},
  journal={Operations Research},
  volume={55},
  number={2},
  pages={191--197},
  year={2007},
  publisher={INFORMS}
}

@article{socialReview06,
  title={Issues in Multiagent Resource Allocation},
  author={Chevaleyre, Y and Dunne, PE and Endriss, U and Lang, J and Lema{\^\i}tre, M and Maudet, N and Padget, J and Phelps, S and Rodriguez-Aguilar, JA and Sousa, P},
  journal={Informatica},
  volume={30},
  pages={3-31},
  year={2006}
}

@ARTICLE{rrr17,
   author ="I. Rodr{\'\i}guez and P. Rabanal and F. Rubio",
   title ="How to make a best-seller: Optimal product design problems",
   journal ="Applied Soft Computing",
   year ="2017",
   volume = "55",
   pages ="178-196"
}

@article{nnr14,
  author    = {N.{-}T. Nguyen and
               T.T. Nguyen and
               M. Roos and
               J. Rothe},
  title     = {Computational complexity and approximability of social welfare optimization
               in multiagent resource allocation},
  journal   = {Autonomous Agents and Multi-Agent Systems},
  volume    = {28},
  number    = {2},
  pages     = {256--289},
  year      = {2014}
}

@article{ems11,
  author    = {U. Endriss and
               N. Maudet and
               F. Sadri and
               F. Toni},
  title     = {Negotiating Socially Optimal Allocations of Resources},
  journal   = {CoRR},
  volume    = {abs/1109.6340},
  year      = {2011}
}

@inproceedings{rr10,
 author = {M. Roos and J. Rothe},
 title = {Complexity of Social Welfare Optimization in Multiagent Resource Allocation},
 booktitle = {AAMAS'10},
 year = {2010},
 isbn = {978-0-9826571-1-9},
 location = {Toronto, Canada},
 pages = {641--648},
}

@article{bd05,
author = {Bez\'{a}kov\'{a}, Ivona and Dani, Varsha},
title = {Allocating Indivisible Goods},
year = {2005},
issue_date = {April 2005},
publisher = {Association for Computing Machinery},
address = {New York, NY, USA},
volume = {5},
number = {3},
journal = {SIGecom Exch.},
month = apr,
pages = {11-18},
numpages = {8}
}

\end{document}